\documentclass[preprint,superscriptaddress]{revtex4-1}
\usepackage{graphics,graphicx,epsfig}

\begin{document}

\title{Evolution of spatially embedded branching trees with interacting nodes}

\author{F. L. Forgerini}
\email{fabricio$\_$forgerini@ufam.edu.br}
\affiliation{Departamento de F\'isica, I3N, Universidade de Aveiro, 3810-193 Aveiro, Portugal}
\affiliation{ISB, Universidade Federal do Amazonas, 69460-000 Coari-AM, Brazil}
\author{N. Crokidakis}
\affiliation{Departamento de F\'isica, I3N, Universidade de Aveiro, 3810-193 Aveiro, Portugal}
\affiliation{Instituto de F\'isica, Universidade Federal Fluminense, 24210-340 Niter\'oi-RJ, Brazil}
\author{S. N. Dorogovtsev}
\affiliation{Departamento de F\'isica, I3N, Universidade de Aveiro, 3810-193 Aveiro, Portugal}
\affiliation{A. F. Ioffe Physico-Technical Institute, 194021 St. Petersburg, Russia}
\author{J. F. F. Mendes}
\affiliation{Departamento de F\'isica, I3N, Universidade de Aveiro, 3810-193 Aveiro, Portugal}

\date{\today}
\begin{abstract}
We study the evolution of branching trees embedded in Euclidean spaces with suppressed branching of spatially close nodes. This cooperative branching
process accounts for the effect of overcrowding of nodes in the embedding space and mimics the evolution of life processes (the so-called ``tree of life'') in which a new level of complexity emerges as a short transition followed by a long period of gradual evolution or even complete extinction. We consider the models of branching trees in which each new node can produce up to two twigs within a unit distance from the node in the Euclidean space, but this branching is suppressed if the newborn node  is closer than at distance $a$ from one of the previous generation nodes. This results in an explosive (exponential) growth in the initial period, and, after some crossover time $t_x \sim \ln(1/a)$ for small $a$, in a slow (power-law) growth. This special point is also a transition from ``small'' to ``large words'' in terms of network science. We show that if the space is restricted, then this evolution may end by extinction. 
\end{abstract}


\maketitle

\section{Introduction}

``The evolution of life is, obviously, a nonuniform process'' \cite{koonin}. For biological evolution, this means that new types of biological objects emerge abruptly with subsequent gradual evolution. This evolutionary process can be schematically depicted as a tree (``the tree of life'' \cite{koonin,moret}), where branches, are, for example, different species. Importantly, the growth of this tree is complicated by interaction and competition between species. In this work we discuss one of the simplest models of growing trees which can mimic this process. 
Growing trees naturally represent a wide range of real-life processes and phenomena \cite{paulo_murilo,sergey1,moret,huson,livro:enc_comp_sys_branching,Cowlishaw,gray,Dorogovtsev:dgm2008,livro_sergey}.  

The Galton-Watson branching process \cite{Kendall} provides a simple example of a growing tree with non-interacting nodes and so uncorrelated branching. 
Interacting branching processes are much more interesting and difficult for analysis \cite{paczuski}. In this work we study evolving trees which evolution is influenced by interaction between some of existing nodes, for example, nodes of the previous generation. We assume that the growing tree is embedded in some metric space and assume that spatially close nodes of the previous generation suppress mutually their ability to born new nodes. In other words, overcrowding of nodes in the embedding space suppresses their ``fertility''. This kind of interaction (competition),  leading to suppression of branching, emerges if there is no sufficient space, no niches for the new nodes and branches (species). 

Due to the embedding space, we can introduce distance between two nodes other then the shortest path internode distance for this tree. For the sake of simplicity, we consider a $D$-dimensional Euclidean space, although the results do not depend qualitatively on $D$. Networks embedded in metric spaces and their evolution already attracted much attention \cite{Dall:dc02,Kleinberg:k00,Carmi:ccs09,Cartozo:cd09,Boguna:bk09,Krioukov:kpk10}. In this paper we are particularly interested in a transition (actually, crossover) between different regimes of the network growth, namely, explosive (exponential) evolution and gradual (power-law) one. Here the evolution of the network is characterized by the variation of the number of its nodes (which corresponds to biological diversity, for example). We find the position of this transition and express it in terms of a single model parameter. This transition coincides with crossover from a ``small word'' to ``large word'' network architectures \cite{Dorogovtsev:dkm08}, where small words show a logarithmic dependence of network diameters on their sizes (total numbers of nodes) and large words show a power-law dependence \cite{Albert:ab02,Newman:n03}. 

One should emphasize a principal difference from the previous studies of this crossover. In Ref.~\cite{Dorogovtsev:dkm08}, the crossover was controlled by a model parameter, while in the present study the small-world and large-world architectures are realized on different stages of the network evolution. 
In addition we find how the spatial distribution of nodes evolves. We also consider the evolving trees embedded in restricted areas of metric spaces, and investigate the possibility of complete extinction. The term extinction for these trees implies the end of evolution and the absence of new generations.

The paper is organized as follows. In Sec.~II we present the model of an evolving tree with interacting nodes and obtain simple estimates for its growth. 
In Sec.~III we describe the network size evolution for two particular models of these trees.  
In Sec.~IV we consider the trees embedded in a bounded space and extinction. In Sec.~V we describe the evolution of the node spatial distribution. Finally, in Sec.~VI, we summarize our results.   


\begin{figure*}
\includegraphics[width=.3\textwidth]{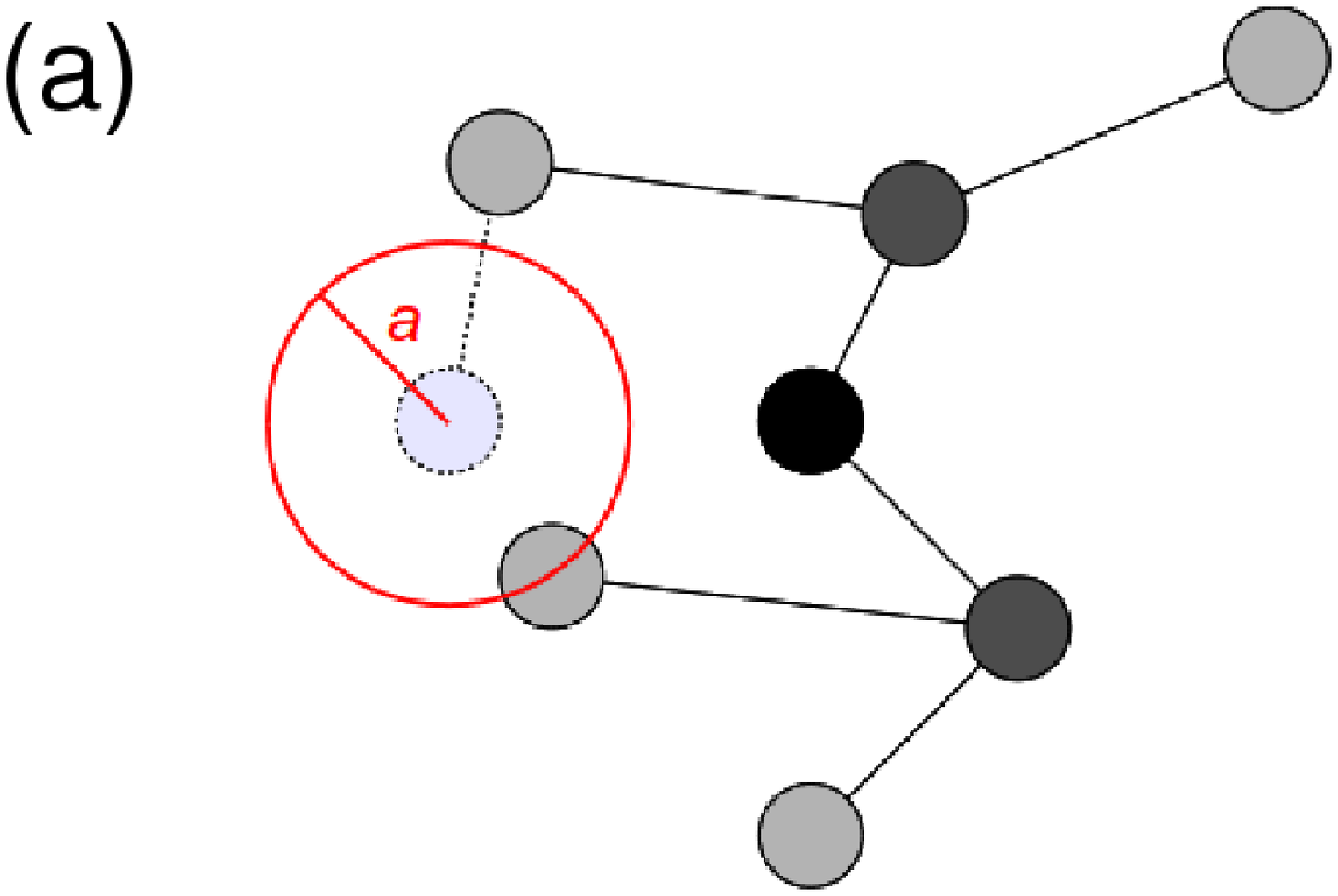}
\includegraphics[width=.3\textwidth]{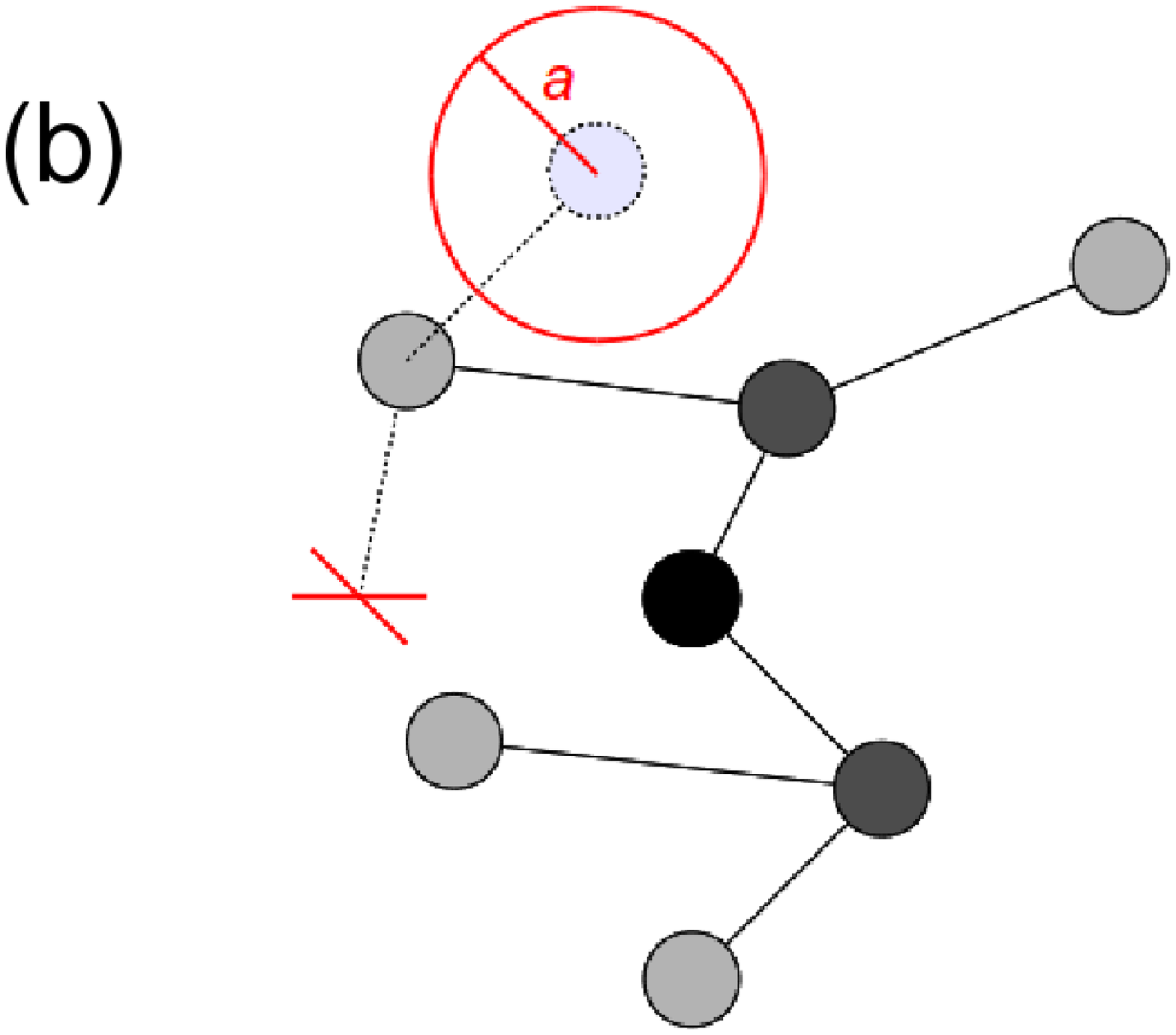}
\includegraphics[width=.3\textwidth]{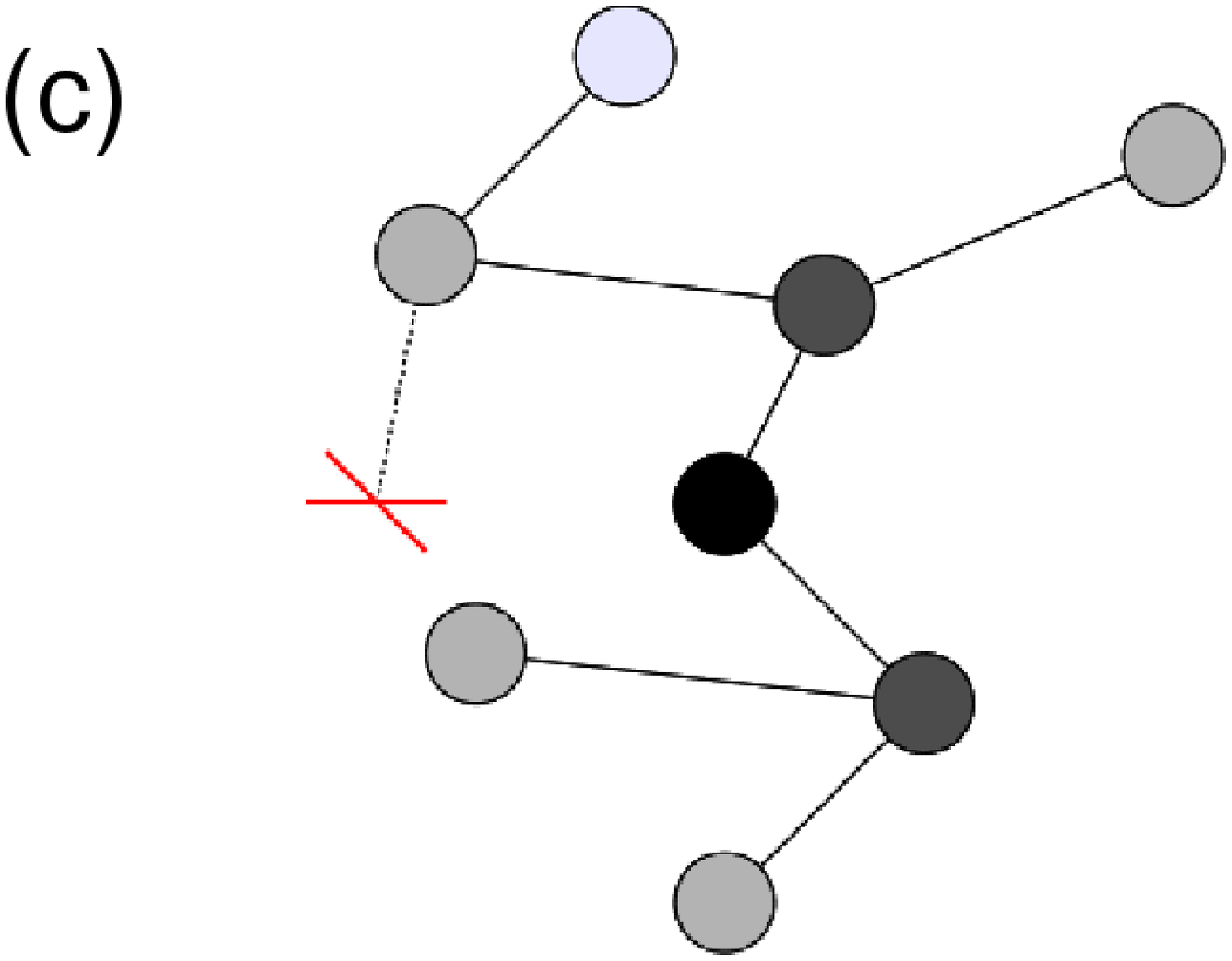}\hfill
\caption{(Color online) 
The scheme of the network growth on a plane. The black node shows the root, the nodes of the first and the second generations are dark and light grey, respectively. The furthermost left node attempts to born two children. The first attempt (a) is abandoned because of the nearby second generation node. The second attempt (b) is successful since the new node has no second generation node within radius $a$ from it. This results in the network (c).  
}
\label{fig1}
\end{figure*}


\section{The Model}
\label{model}

The model, which we use, is formulated as follows (see Fig.~\ref{fig1} showing schematically the grows of the tree embedded in a two-dimensional space). The growth of the tree starts from a root node. At each time step, each of the nodes of the tree attempts to emit two leaves (leaf is a link with a new node),  so at each time step a new generation of nodes is born. The network is embedded in a $D$-dimensional Euclidean space, and the root has zero coordinates. At each time step, make the following:

\begin{enumerate}    

\item[(1)] Choose uniformly at random a node $i$ (coordinates ${\bf x}_i$) from the previous generation 
and make an attempt to create its leaf with a new node at the point ${\bf x}_i+{\bf \Delta}_i$. Here the random vector ${\bf \Delta}_i$ is uniformly distributed within $-1{\leq}\Delta_{x,i}{\leq}1$, $-1{\leq}\Delta_{y,i}{\leq}1$, \ldots . If among the nodes of the previous generation (excluding the parent node $i$) and among the nodes already created at this time step, no nodes are closer than at distance $a$ from the point, ${\bf x}_i+{\bf \Delta}_i$, then create the leaf. If such nodes exist, abandon this attempt. Make the next attempt to create the second leaf from this node using the same rules. 

\item[(2)] From the rest nodes of the previous generation, choose uniformly at random nodes one by one and repeat (1) again and again until all the nodes of the previous generation will be updated.  

\end{enumerate}

We will also consider a variation of this model, in which for each attempted node birth, closeness to all existing nodes should be checked and not only to the previous generation nodes. Importantly, in both these models, existing nodes and links never disappear. 

\begin{figure}
 \includegraphics[width=0.7\linewidth]{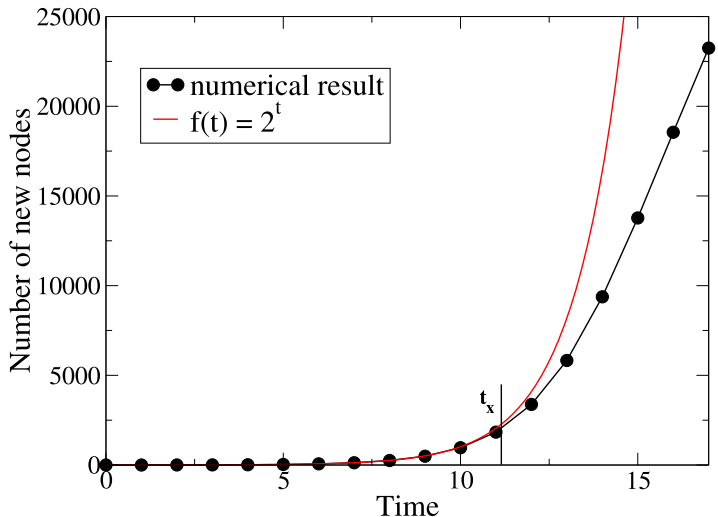}
 \caption{(Color online) The number of new nodes in the tree as a function of time, $D = 1$ and $a = 0.1$.}
 \label{example}
\end{figure}

Figure~\ref{example} shows the result of simulation of this model for $D=1$ and sufficiently small $a$, namely, the evolution of the number of nodes $N(t)$ of generation $t$ which plays the role of time. Initially, $N$ grows exponentially, $N=2^t$. One can see that after certain crossover time $t_x$, the network growth is slower than exponential. For an arbitrary dimension $D$, one can easily estimate $N \cong \text{const\,}(t/a)^D$ at large $t$. 
To obtain this estimate, we assume that nodes of generation $t$ are within a hypersphere which radius grows with a constant rate of the order of $1$ (the rate is actually smaller than $1$). This average rate of expansion is explained by the fact that children in this tree are born within unit distance from their parent nodes. Since the neighboring nodes cannot be closer than at distance $a$, we obtain $N \sim t^D/a^D$. 

Note that if the parameter $a$ is sufficiently large, $N$ does not grow at all. If $a$ is, say $2$, $N=1$ for any $t$, and our tree is a chain. 

From $N(t\lesssim t_x)=2^t$ and $N(t\gtrsim t_x)\sim (t/a)^D$, we have very roughly 

\begin{equation}
2^{t_x} \sim (t_x/a)^D ,
\label{e1}
\end{equation}
which leads to the estimate

\begin{equation}
t_x \sim \frac{D}{\ln2}\ln\Bigl(\frac{1}{a}\Bigr) ,
\label{e2}
\end{equation}
at small $a$. In Sec.~\ref{crossover_time}, we will demonstrate that this simple estimate describes well the results of our simulations. 


\section{Network size evolution} 
\label{size}

\begin{figure}
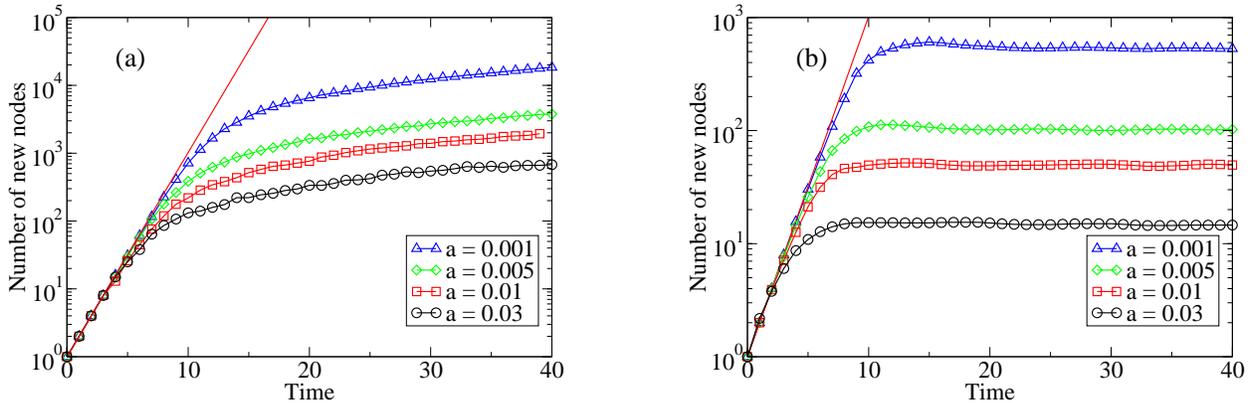

\vspace{10pt}
    \begin{minipage}[t]{0.45\linewidth}
	\includegraphics[width=1\linewidth]{bio_nets_fig3a.eps}\\
    \end{minipage}\hfill
    \begin{minipage}[t]{0.45\linewidth}
	\includegraphics[width=1\linewidth]{bio_nets_fig3b.eps}\\
    \end{minipage}\hfill
\caption{(Color online) Evolution of the number of new nodes in the networks for different values of the parameter $a$ ($D=1$). The data were obtained after averaging over 100 samples. (a) The trees evolve according to the rules introduced in Sec.~\ref{model}, i.e., only the previous generation nodes influence the branching process. (b) The trees in which newborns cannot be closer than at distance $a$ from any of existing nodes (apart of their parents), i.e. the branching process is influenced by all existing nodes.}
 \label{new_nodes_X_time}

\end{figure}

\begin{figure}[!htb]
 \includegraphics[width=0.7\linewidth]{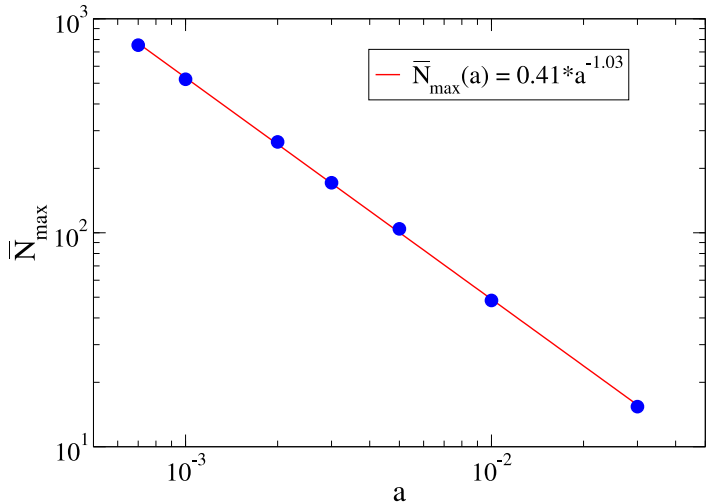}
 \caption{(Color online) Log-log plot of $\overline{N}_{max}$ versus $a$ obtained by simulating the model of trees in which newborn nodes cannot be closer than at distance $a$ from any of existing nodes (apart of their parents). The straight line has slope -1.}
 \label{nmax_all}
\end{figure}

Data similar to Fig.~\ref{example} are shown on the normal-log plot, Fig.~\ref{new_nodes_X_time}(a), for a few values of $a$ ($D=1$). The straight line in the figure is the dependence $2^t$, and the crossover from the exponential to a slower growth is clearly seen. Figure~\ref{new_nodes_X_time}(a) was obtained from the model formulated in Sec.~\ref{model}, in which the previous generation nodes affect the branching process. We performed similar simulations for the model, in which newborn nodes cannot be closer than at distance $a$ from any of existing nodes (apart of their parents). The results of the simulations (the evolution of the number of nodes of generation $t$) are shown in Fig.~\ref{new_nodes_X_time}(b). In contrast to Fig.~\ref{new_nodes_X_time}(a), in the network in which all nodes influence branching, the number of nodes of generation $t$ approaches a constant value ${\overline N}_{max}(a)$ at large $t$. One can easily obtain this plato using an estimate similar to that from the previous section, $N_{tot} \cong \text{const\,}(t/a)^D$. The only difference is that now $N_{tot}$ in that estimate is the total number of nodes in the network, and so for the number of nodes of generation $t$, we have $N_t = dN_{tot}(t)/dt \sim D t^{D-1}/a^D$, and, in particular, $N_t\sim 1/a$ at $D=1$. The results of simulations for this model, which give ${\overline N}_{max}\approx0.4/a$, see Fig.~\ref{nmax_all}, agree with this simple estimate.


\begin{figure}
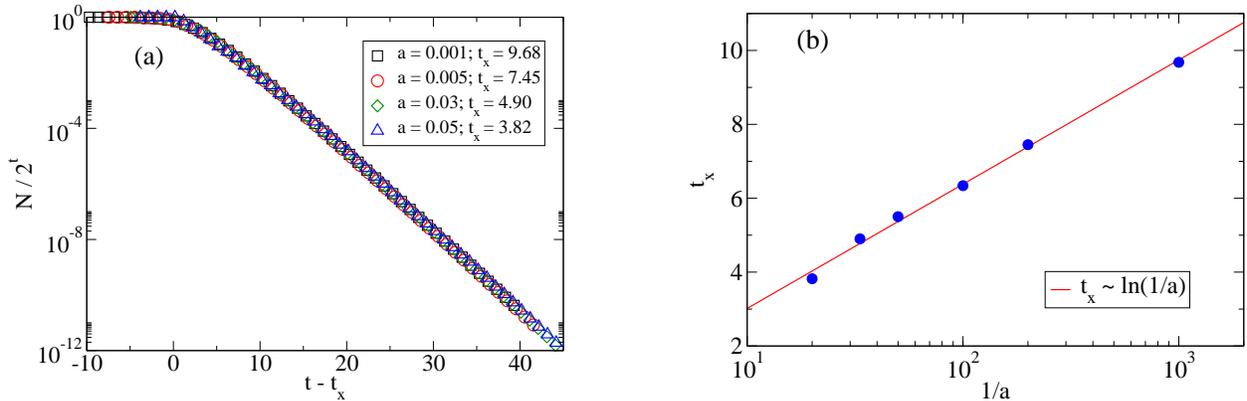

\vspace{10pt}
    \begin{minipage}[t]{0.45\linewidth}
	\includegraphics[width=1\linewidth]{bio_nets_fig5az.eps}\\
    \end{minipage}\hfill
    \begin{minipage}[t]{0.45\linewidth}
	\includegraphics[width=1\linewidth]{bio_nets_fig5bz.eps}\\
    \end{minipage}\hfill
 \caption{(Color online) Finding the crossover time $t_x$ from simulation data for the model from the previous section. $N\equiv N_t$ is the number of the $t$-generation nodes. The dependencies $N/2^t$ versus $t - t_x$ for different values of $a$~and $D=1$ (a) collapse into a single curve for the crossover times $t_x(a)$ shown on panel (b). Fitting gives $t_x(a) = -0.34 + 1.46\ln(1/a)$, which is consistent with Eq.~(\protect\ref{e2}).}
 \label{reescal_tx}
\end{figure}

The crossover time between two regimes of the network evolution is obtained in Fig.~\ref{reescal_tx} for the growing tree model ($D=1$) from the previous section. Note that the result, $t_x(a) = -0.34 + 1.46\ln(1/a)$, agrees well with Eq.~(\ref{e2}), $t_x \approx (D/\ln 2)\ln(1/a)$, since $1/\ln2=1.44...$. Clearly, the time $t$ is of the order of the diameter $d$ of this tree (the maximum separation between two nodes in a network). So we have the logarithmic dependence of the diameter $d$ on the total number $N_{tot}$ of nodes in these trees for $t\ll t_x$, and the power-law dependence $d(N_{tot})$ for $t\gg t_x$, which corresponds, respectively, to the small-world and large-world network architectures.


\section{Spatial restriction}
\label{restriction}

It is natural to introduce a spatial restriction into the models.  
If our network is embedded in an infinite Euclidean space, the evolution is actually determined by the only parameter $a$ (recall that we set the scale of displacements of children nodes from their parents to $1$, i.e., this is the unit distance in this problem). If the area of the space, in which the network is embedded, is restricted, then the model has an extra parameter, namely the linear size of the area, $L$. For simplicity, we assume that this area does not change with time. 

Let the Euclidean coordinates ${\bf x}_i$ of all nodes in the network be within the area $-L<x_i<L$, $-L<y_i<L$, \ldots . In our simulations we use
periodical boundary conditions, but, in principle, this is not necessary. If $L$ is finite, then one may expect that the size of the tree will finally approach some limiting value. The network has even a chance to extinct if at some moment all its nodes occur in one small area. 
In, e.g., population biology, the smaller a population, the more susceptible it is to extinction by various causes \cite{Sznajd-Weron}.

\begin{figure}
 \includegraphics[width=0.7\linewidth]{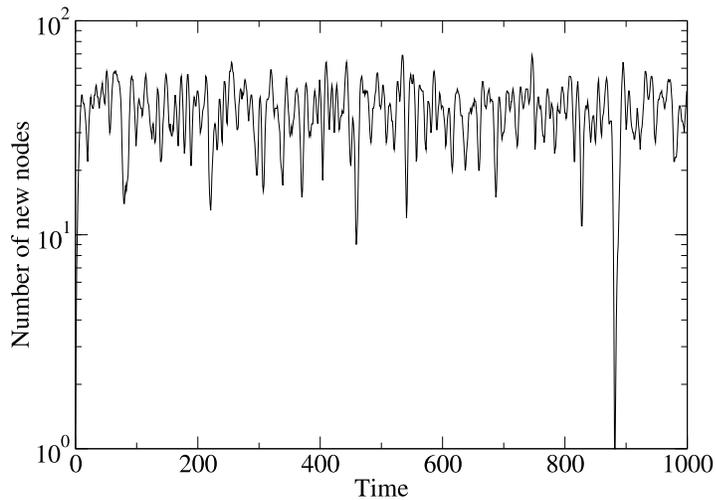}
 \caption{ Evolution of the number of new nodes (number of nodes in the generation $t$) for a single realization of the network of Sec.~\protect\ref{model} defined on a one-dimensional interval $-L<x<L$, where $L=1$, $a=0.1$.}
 \label{spatial_restriction}
\end{figure}

Figure~\ref{spatial_restriction} demonstrates an example of the evolution of the network, which in this case has $a=0.1$ and $L=1$. The network rapidly enters the fluctuation regime, in which $N_t$ fluctuates around a mean value $\overline{N}_{max}$, and extincts before 900 time steps. After that we again introduced a root and restarted the process.   
The picture which we observe agrees with traditional views on extinction processes which show  ``relatively long periods of stability alternating with short-lived extinction events'' (D.~M.~Raup) \cite{raup}. This kind of extinction may occur in branching annihilating random walks and other related processes studied in Refs.~\cite{takayasu, jensen}. In other models of biological evolution, extinction may require external factors or an environmental stress \cite{newman} or an internal mechanism, such as a mutation \cite{sibani}.

\begin{figure}
 \includegraphics[width=0.7\linewidth]{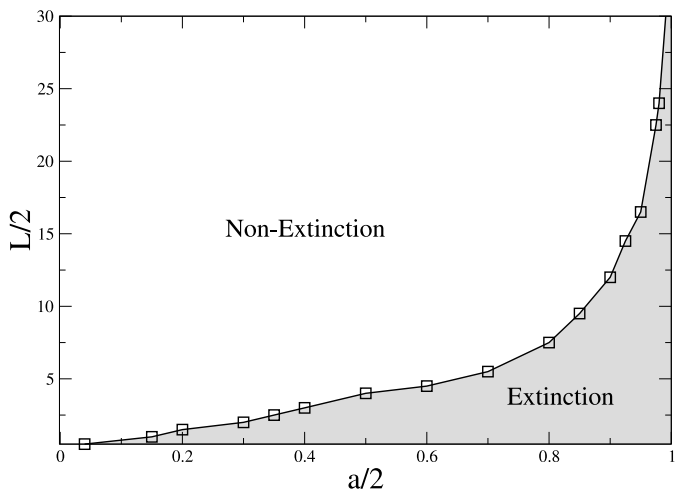}
 \caption{Extinction of the network embedded in the $(-L,L)$ ring during $10^5$ generations. The extinction and non-extinction regions are present on the $L/2$ vs. $a/2$ diagram.}
 \label{phase_diagram}
\end{figure}

For the same model, we investigated the state of the branching process after $t_{\text{observation}}= 10^5$ generations (i.e., time steps) for various $L>1$ and $a<2$ (for $a>2$, the network turns out to be a chain). In other words, we analysed if the extinction time for given $L$ and $a$ is smaller than $10^5$ generations or not. On the $(a/2,L/2)$ diagram, Fig.~\ref{phase_diagram}, the boundary separating the extinction and non-extinction regions is a monotonously growing curve $L(a)$, where $L(a) \propto a$ at sufficiently small $a$ and diverges as $a$ approaches $2$. Note that $L(a)$ actually depends on $t_{\text{observation}}$, and with increasing observation time, the area of extinction should increase. 
Based on simulations, it is practically impossible to check whether $L(a,t_{\text{observation}}{\to}\infty)$ approachs some non-trivial limit or, for any finite $L$ and non-zero $a$, the network should finally extinct. 
The latter option seems to be more plausible.


\begin{figure}
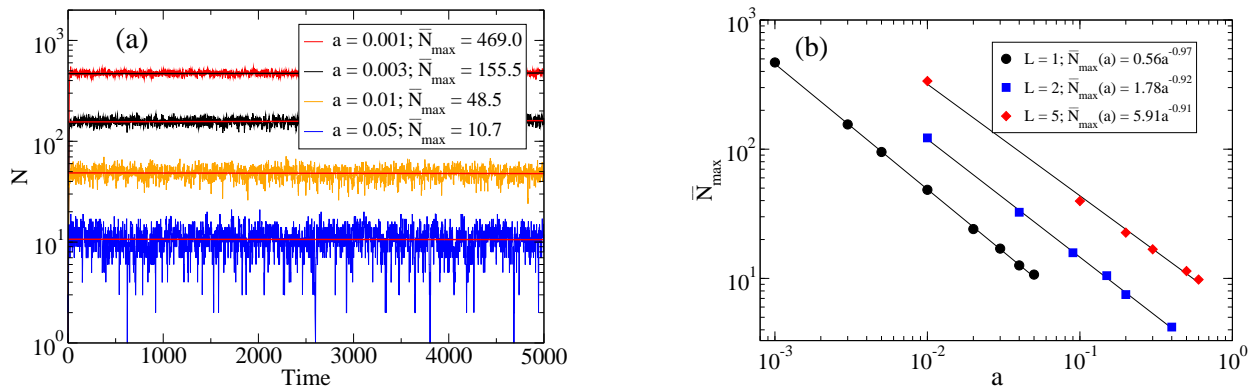

\vspace{20pt}
    \begin{minipage}[t]{0.45\linewidth}
	\includegraphics[width=1\linewidth]{bio_nets_fig8az.eps}\\
    \end{minipage}\hfill
    \begin{minipage}[t]{0.45\linewidth}
	\includegraphics[width=1\linewidth]{bio_nets_fig8bz.eps}\\
    \end{minipage}\hfill
\caption{(Color online) (a) Variation of the number of nodes $N_t$ in the current generation with time, for different values of $a$. The network is embedded in the interval $-L \leq x \leq L$, where $L=1$, and only the previous generation influence the branching process. The average value of $N_t$ at large $t$, $\overline{N}_{max}$, is represented by a solid straight line. (b) $\overline{N}_{max}$ versus $a$ for different $L$.}
 \label{N_a}
\end{figure}

Figure~\ref{N_a}(a) shows the evolution of $N_t$ for a few different values of $a$. The averaged $N_t$ (averaged over times before extinction) $\overline{N}_{max}$, decreases with $a$ as Fig.~\ref{N_a}(b) demonstrates.  
The simplest estimation give $\overline{N}_{max}(a,L) \sim L/a$. Figure~\ref{N_a}(b) confirms that this estimate is reasonable, $\overline{N}_{max}$ indeed inversely proportional to $a$, although these simulations indicate deviation from proportionality on $L$ for sufficiently large $L$. 
Since new nodes are born uniformly randomly in the interval $(-1,1)$ from their parents, the case of $L=1$ is special. In this situation, new nodes are actually born at any point of the ring with equal probability independently of the positions of their parents, and so a network structure here is not essential. One can consider this specific model with new nodes born in arbitrary points with equal probability at arbitrary $L$ and find $\overline{N}_{max}(a,L) \approx 0.5 L/a$. Figure~\ref{N_a}(b) for our original model shows a functionally faster growth of $\overline{N}_{max}(a,L)$ with $L$ than this proportional dependence.    
Note finally that the deviations of fluctuating $N_t$ from the mean values $\overline{N}_{max}$ in Fig.~\ref{N_a}(a) are of the order of $\sqrt{\overline{N}_{max}}$ for each $L$ and $a$.

\begin{figure*}
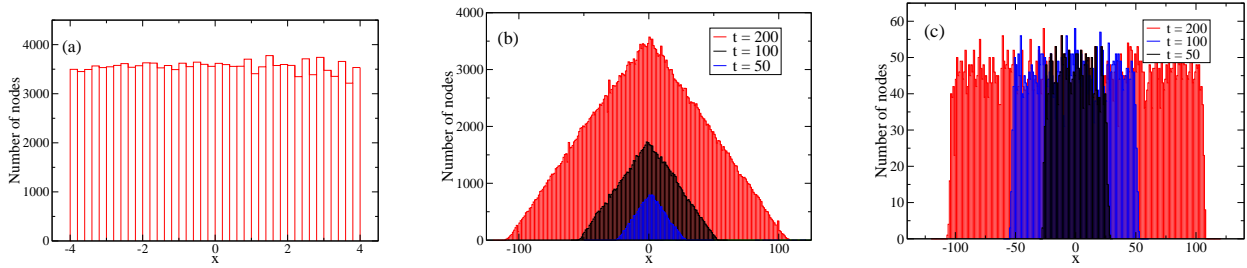

\vspace{0.25cm}
\includegraphics[width=.3\textwidth]{bio_nets_fig9a.eps}\hfill
\includegraphics[width=.3\textwidth]{bio_nets_fig9b.eps}\hfill
\includegraphics[width=.3\textwidth]{bio_nets_fig9cp.eps}\hfill
\caption{(Color online) 
Distribution of nodes of the growing trees in space. (a) The node spatial distribution of the tree embedded in the interval $-1 \le x \le +1$ after 1000 time steps. The birth of new nodes in the tree is influenced only by a previous generation. The vertical columns of the histogram show the numbers of nodes within bins of width $0.2$. (b) The node spatial distributions for the tree embedded in a one-dimensional space at different instants of the growth. The birth of new nodes in the tree is influenced only by a previous generation. The vertical columns of the histogram show the numbers of nodes within bins of width $1$. (c) The same as for (b), but the birth of new nodes in the tree is influenced by all existing nodes. For all three plots, $a = 0.01$. Each of the results was obtained from a single realization.}
\label{distribution}
\end{figure*}


\section{Node spatial distribution}
\label{crossover_time}

In general, the nodes of the growing trees under consideration are non-uniformly distributed in the embedding spaces. Only if the embedding area is restricted, the spatial distribution finally becomes uniform, see Fig.~\ref{distribution}(a). For infinite embedding space, the evolution of the node spatial distributions is shown in Figs.~\ref{distribution}(b) and (c) for the trees in which the birth of new nodes is determined only by a previous generation and by all existing nodes, respectively. The distributions in three instances are shown. The triangular shape of these distributions in Fig.~\ref{distribution}(b) indicate that the spatial distribution of nodes of generation $t$ has a symmetric step-function form with boarders moving away from the center (root) with constant velocity equal approximately to $0.5$, so that their coordinates increase proportionally to $t$. The density of nodes between the borders is a constant equal approximately to $0.2/a$. In the second case, Fig.~\ref{distribution}(c), this expanding step-function form describes the evolution of the spatial distribution of all nodes in the tree. The border speed is approximately $0.6$, and the density of nodes between borders is a constant equal approximately to $0.45/a$. (Note that, as it should be, this value is close to the number $\overline{N}_{max}\approx 0.4/a$ of new nodes found for the this tree in Sec.~\ref{size}, see Fig.~\ref{nmax_all}.) These observations explain the high quality of simple estimates obtained in Sec.~\ref{model}. 

Finally, for the networks embedded in a restricted area, in which the birth of new nodes is determined by a previous generation, we also measured the distribution of the number of nodes in one generation. We observed that this distribution is centered at $\overline{N}_{max}$ and is close to the normal distribution.


\section{Conclusions}

We have studied the models of evolving trees embedded in a Euclidean space, in which the branching process is determined by the relative position of nodes in space. In these models, overcrowding suppresses the ``fertility'' of nodes. We have investigated two regimes of the evolution of these trees and crossover between them. In the initial stage of evolution, the network growth is exponentially fast, and the network is a small world. After some crossover time, this network becomes to grow much slowly, and, in this regime, the network has a large world architecture in terms of network science. We have demonstrated that the embedding of the network in a restricted area, which is natural for general evolution, sets limits to growth and can result in complete extinction. The simplest models which we analysed can only schematically describe real evolution processes in biology. Even these null models however are sufficient to demonstrate the transition from an explosive to gradual evolution accompanied by a dramatic change of the network structure. We believe that the significance of the network representation of evolutionary processes, e.g., the so-called ``tree of life'', is greater than simply being 
a convenient visualization. We suggest that thorough exploration of the structural organization of the empirical trees of life and their analogies on different stages of evolution will essentially improve our understanding of evolutionary processes.

\begin{acknowledgements} 
This work was partially supported by projects  
PTDC/FIS/108476/2008, PTDC/SAU-NEU/103904/2008, and PTDC/MAT/114515/2009.  
N. Crokidakis acknowledges financial support from the Brazilian agencies CNPq and CAPES. F.~L. Forgerini was supported by FCT project SFRH/BD/68813/2010.
\end{acknowledgements}


\begin{thebibliography}{16}

\bibitem{koonin}
E. V. Koonin, Biology Direct. \textbf{2}, 21 (2007).

\bibitem{moret}
B. M. E. Moret, L. Nakhleh, T. Warnow, C. R. Linder, A. Tholse, A. Padolina, J. Sun, and R. Timme,
IEEE ACM Trans. Comput. Biol. Bioinf. \textbf{1}, 13 (2004).

\bibitem{paulo_murilo}
V. Schw\"ammle and P. M. C. de Oliveira,
Physica A \textbf{388}, 2874 (2009).

\bibitem{sergey1}
S. N. Dorogovtsev and J. F. F. Mendes,
Proc. R. Soc. London B \textbf{268}, 2603 (2001).

\bibitem{huson}
D. H. Huson and D. Bryant,
Mol. Biol. Evol. \textbf{23}, 254 (2006).

\bibitem{livro:enc_comp_sys_branching}
M. J. Alava and K. B. Lauritsen,
in {\textit{Encyclopedia of Complexity and Systems Science}}, edited by B. Meyers. Springer, Heidelberg, 2009, p.~644.

\bibitem{Cowlishaw}
G. Cowlishaw and R. Mace,
Ethol. Sociobiol. \textbf{17}, 87 (1996).

\bibitem{gray}
R. D. Gray and Q. D. Atkinson, 
Nature \textbf{426}, 435 (2003). 

\bibitem{Dorogovtsev:dgm2008}
S.~N. Dorogovtsev and J.~F.~F. Mendes, Adv. Phys. \textbf{51}, 1079 (2002); S. N. Dorogovtsev, A. V. Goltsev, and J.~F.~F.~Mendes, 
Rev. Mod. Phys. \textbf{80}, 1275 (2008). 

\bibitem{livro_sergey}
S.~N. Dorogovtsev and J.~F.~F. Mendes, \textit{Evolution of Networks} (Oxford University Press, Oxford, 2003); 
S.~N. Dorogovtsev, \textit{Lectures on Complex Networks} (Oxford University Press, Oxford, 2010).

\bibitem{Kendall}
D. G. Kendall
J. London Math. Soc. \textbf{s1-41}, 385 (1966).

\bibitem{paczuski}
V. Sood, M. Mathieu, A. Shreim, P. Grassberger, and M. Paczuski, 
Phys. Rev. Lett. \textbf{105}, 178701 (2010). 

\bibitem{Dall:dc02}
J.~Dall and M.~Christensen, 
Phys. Rev. E \textbf{66}, 016121 (2002). 
 
\bibitem{Kleinberg:k00}
J. Kleinberg, 
Nature \textbf{406}, 845 (2000). 

\bibitem{Carmi:ccs09}
S.~Carmi, S.~Carter, J.~Sun, and D. ben-Avraham, 
Phys. Rev. Lett. \textbf{102}, 238702 (2009). 

\bibitem{Cartozo:cd09}
C.~C. Cartozo and P. De~Los~Rios,  
Phys. Rev. Lett. \textbf{102}, 238703 (2009). 

\bibitem{Boguna:bk09}
M.~Bogu\~{n}\'a and D.~Krioukov,   
Phys. Rev. Lett. \textbf{102}, 058701 (2009).  

\bibitem{Krioukov:kpk10}
D.~Krioukov, F.~Papadopoulos, M.~Kitsak, A. Vahdat, M. Bogu\~{n}\'a, 
Phys. Rev. E \textbf{82}, 036106 (2010).  

\bibitem{Dorogovtsev:dkm08}
S.~N.~Dorogovtsev, P.~L.~Krapivsky, and J.~F.~F. Mendes, 
EPL \textbf{81}, 30004 (2008). 

\bibitem{Albert:ab02}
R.~Albert and A.-L.~Barab\'asi,  
Rev. Mod. Phys. \textbf{74}, 47 (2002). 

\bibitem{Newman:n03}
M.~E.~J. Newman,
SIAM Review \textbf{45}, 167 (2003). 

\bibitem{Sznajd-Weron}
K. Sznajd-Weron and R. Weron, 
Physica A \textbf{293}, 559 (2001).

\bibitem{raup}
M. D. Raup, 
Science \textbf{231}, 4745 (1986).

\bibitem{takayasu}
H. Takayasu and A.~Y. Tretyakov,
Phys. Rev. Lett. \textbf{68}, 3060 (1992).

\bibitem{jensen}
I. Jensen,
Phys. Rev. E \textbf{50}, 3623 (1994). 

\bibitem{newman}
B.~W. Roberts and M. E. J. Newman, 
J. Theor. Biol. \textbf{180}, 39 (1996).

\bibitem{sibani}
P. Sibani, M. R. Schmidt, and P. Alstr\o{}m,
Phys. Rev. Lett. \textbf{75}, 2055 (1995).


\end{thebibliography}
\end{document}